\documentclass[12pt,a4paper]{article}

\usepackage{graphicx}

\begin{document}

\title{Multi-Fluid Simulation of the Magnetic Field Evolution in Neutron Stars}

\maketitle

\begin{center}
Jaime Hoyos$^*$, Andreas Reisenegger$^*$, Juan A. Valdivia$^+$
\end{center}

\begin{center}
$^*$Depto. de Astronomia y Astrofisica, Pontificia Universidad Catolica de Chile,Casilla 306, Santiago 22, Chile
\end{center}
\begin{center}
$^+$Depto. de Fisica, Facultad de Ciencias, Universidad de Chile, Casilla 653, Santiago, Chile
\end{center}

\begin{center}
\textbf{Abstract}
\end{center}
Using a numerical simulation, we study the effects of \emph{ambipolar diffusion} and \emph{ohmic diffusion} on the magnetic field evolution in the interior of an isolated neutron star. We are interested in the behavior of the magnetic field on a long time scale, over which all Alfv\'en and sound waves have been damped. We model the stellar interior as an electrically neutral plasma composed of neutrons, protons and electrons, which can interact with each other through collisions and electromagnetic forces. Weak interactions convert neutrons and charged particles into each other, erasing chemical imbalances. As a first step, we assume that the magnetic field points in one fixed Cartesian direction but can vary along an orthogonal direction. We start with a uniform-density background threaded by a homogeneous magnetic field and study the evolution of a magnetic perturbation as well as the density fluctuations it induces in the particles. We show that the system evolves through different quasi-equilibrium states and estimate the characteristic time scales on which these quasi-equilibria occur.

\section{Introduction}

It is thought that the decay of the magnetic field in \emph{magnetars} is the main source of its X-ray and $\gamma$-ray luminosity since these objects appear to radiate substantially more power than available from the energy rotational loss \cite{TD-96}. The spontaneous decay of the magnetic field could occur through \emph{ambipolar diffusion}, \emph{Hall drift} and \emph{ohmic diffusion}, which are non-ideal magnetohydrodynamics processes that occur in thousands of years, compared to dynamical sound and Alfv\'en time scales of milliseconds to seconds. Ambipolar diffusion promote a dissipative magnetic field advection, through the movement of a bulk of charged particles relative to the neutrons, the \emph{Hall drift} is a non-dissipative advection of the magnetic field caused by the electrical current associated with it, and the magnetic field ohmic diffusion is a dissipative process caused by the electrical resistivity. The time scales of these process were estimated in Ref.~2. For classical pulsar magnetic field strengths, these were found to be longer than the lifetimes of these stars, and therefore unlikely to be observationally important. For the case of magnetars, it was found that magnetic field decay by these processes might be occurring \cite{TD-96,ACT-04}. However, a full understanding of these processes, their interactions, and their effectiveness in neutron stars is still lacking. The work of \cite{GR-92} was analytical, and therefore useful to identify general processes and identify the relevant time scales, but not to address the action of the identified processes in their full nonlinear development, and their interaction with each other. The full evolution of the magnetic field can only be addressed by numerical simulations. Recent ideal three-dimensional single-fluid magnetohydrodynamics simulations showed that, in a stably stratified star, a complicated, random, initial field generally evolves on a short, Alfv\'en-like time scale to a relatively simple, roughly axisymmetric, large-scale configuration containing a toroidal and a poloidal component of comparable strength, both of which are required in order to stabilize each other \cite{BS-04}. It is interesting to study the effects of the non-ideal processes studied in Ref.~2 on the evolution of this configuration. As a first step in this direction, here we simulate the decay of a magnetic field, including the effects studied in  Ref.~2, in a system where the magnetic field points in one Cartesian direction but varies only along an orthogonal direction. It is shown that the magnetic field evolves through different quasi-equilibrium states and we estimate the characteristic time scales where these quasi-equilibria occur.
\section{General Physical Model}
We model the stellar interior as an electrically neutral and lightly ionized plasma composed of three moving particle species: neutrons $(n)$, protons $(p)$, and electrons $(e)$. This is a slight extension of the model of \cite{GR-92}, since these authors considered the neutrons to form a motionless background in diffusive equilibrium. We account for strong interactions between neutrons and protons by writing their chemical potentials as functions of both their number densities $\mu_i(\vec{r},t)=\mu_i(\{n_j(\vec{r},t)\})$, with $i$, $j=n,p$ \cite{A-98}. We consider the electrons as non-interacting particles, thus we write their chemical potential as a function of their own density, $\mu_e(\vec{r},t)=\mu_e(n_e(\vec{r},t))$. The processes of \cite{GR-92} occur in thousands of years, compared to dynamical time scales of milliseconds to seconds, thus we use a slow-motion approximation in which we neglect the inertial terms in the equations of motion for the particles. The plasma species are described as three fluids coupled by collisions and electromagnetic forces, satisfying the equations of motion \cite{GR-92,R-05}:

\begin{eqnarray}
\label{movecond}
\vec{0}&=&-n_i\vec{\nabla}\mu_i-n_i\frac{\mu_i}{c^2}\vec{\nabla \psi}+n_iq_i(\vec{E}+\frac{\vec{v}_i}{c}\times \vec{B})\nonumber\\&&-\sum_{j\neq i}\gamma_{ij}n_in_j(\vec{v}_i-\vec{v}_j).
\end{eqnarray}

In Eq.~\ref{movecond},  $\vec{v}_i$ is the mean velocity, $\mu_i/c^2$ is the effective inertia of each species, $\psi$ is the gravitational potential, $\vec{E}$ and $\vec{B}$ are the electric and magnetic fields, and the last term represents the drag forces due to elastic binary collisions, which damp the relative motions between the different species. The collisional coupling strengths are parametrized by the symmetric matrix $\gamma_{ij}$, whose elements  are inversely proportional to the characteristic collisional times between particles of species $i$ against species $j$ and generally depend on the local density and temperature. Neglecting the acceleration terms in Eq.~\ref{movecond} leads to a set of differential equations that are first, rather than second order in time \cite{GR-92,R-05}. We introduce a slight modification in the equation of motion for the neutrons by introducing an artificial friction-like term of the form $-n_n\alpha \vec{v}_n$. The basic role of this extra term is to allow a fluid element (containing all the species) to relax to the MHD equilibrium in a time scale that should be short compared with the time scales we are interested in, but long compared to the dynamical time-scales,         

\begin{equation}
\label{move2}
\vec{0}=-n_n\vec{\nabla}\mu_n-n_n\frac{\mu_n}{c^2}\vec{\nabla \psi}-\sum_{j\neq n}\gamma_{nj}n_nn_j(\vec{v}_n-\vec{v}_j)-n_n\alpha \vec{v}_n.
\end{equation}

We combine Eq.~\ref{movecond} for the charged particles, with the induction equation $\vec{\nabla}\times \vec{E}=-(1/c)(\partial \vec{B}/\partial t)$, to obtain

\begin{eqnarray}
\label{magn}
\frac{\partial \vec{B}}{\partial t}&=&\vec{\nabla}\times \left[(\vec{v}_n+\vec{v}_A+\vec{v}_H)\times \vec{B}\right]-\vec{\nabla}\times \left(\frac{c^2\vec{\nabla}\times{\vec{B}}}{4\pi\sigma}\right)\nonumber\\&&
-\frac{c}{2e}\vec{\nabla}\left(\frac{\gamma_{en}-\gamma_{pn}}{\gamma_{en}+\gamma_{pn}}\right)\times\vec{\nabla}(\mu_p+\mu_e)\nonumber\\&&
-{1\over ec}\nabla\left({\gamma_{en}\mu_p-\gamma_{pn}\mu_e}\over{\gamma_{en}+\gamma_{pn}}\right)\times\nabla\psi. 
\end{eqnarray}

The first term on the right-hand side of Eq.\ref{magn} shows that the magnetic field is transported by the sum of the neutron bulk velocity $\vec{v}_n$, the ambipolar diffusion velocity $\vec{v}_A$ and the Hall drift velocity $\vec{v}_H$. The second term represents the ohmic diffusion, where the electrical conductivity is 

\begin{equation}
\sigma=\frac{e^2}{\gamma_{ep}+\frac{n_n/n_c}{\frac{1}{\gamma_{pn}}+\frac{1}{\gamma_{en}}}}.
\end{equation}

Finally, the last two terms in right-hand side of Eq.\ref{magn} represent battery terms. The ambipolar and Hall drift velocities are given by the equations:

\begin{eqnarray}
\label{ambi}
\vec{v}_A&=&\frac{\gamma_{pn}(\vec{v}_p-\vec{v}_n)+\gamma_{en}(\vec{v}_e-\vec{v}_n)}{\gamma_{en}+\gamma_{pn}}\nonumber\\&&
=\frac{\left(\vec{j}\times\vec{B}/c\right)-n_c\left(\vec{\nabla}\mu_c+(\mu_c/c^2)\vec{\nabla}\psi\right)}{n_n n_c(\gamma_{en}+\gamma_{pn})},
\end{eqnarray}

\begin{equation}
\label{Hall}
\vec{v}_H=\frac{\gamma_{en}-\gamma_{pn}}{\gamma_{en}+\gamma_{pn}}(\vec{v}_p-\vec{v}_e)=\frac{\gamma_{en}-\gamma_{pn}}{\gamma_{en}+\gamma_{pn}}\frac{c\vec{\nabla}\times\vec{B}}{4\pi n_c e}.
\end{equation}

In Eq.~\ref{ambi}, we defined $\mu_c\equiv \mu_e+\mu_p$, whereas $\Delta \mu= \mu_c-\mu_n$ is the chemical imbalance. Weak interactions between the particles balance chemical potential imbalances between the charged particles and neutrons. We define the difference between the rates, per unit volume, of the reactions $p+e\rightarrow n+\nu_e$ and $n\rightarrow p+e+\bar{\nu_e}$ as $\Delta \Gamma \equiv\Gamma(p+e\rightarrow n+\nu_e)-\Gamma(n\rightarrow p+e+\bar{\nu_e})=\lambda \Delta \mu$, where $\lambda$ generally depends on density and temperature \cite{GR-92}. The continuity equations for the particles are given by the equations:

\begin{equation}
\label{densn}
\frac{\partial n_i}{\partial t}+\vec{\nabla}\cdot(n_i\vec{v_i})=\pm\lambda(\Delta \mu).
\end{equation}

In Eq.~\ref{densn}, the $+$ sign corresponds to the neutrons and the $-$ sign to the electrons and protons. The electrical neutrality condition is $n_e=n_p=n_c$, and we define the \emph{baryon} number density as $n_B=n_n+n_c$. If we add Eq.~\ref{densn} for charged particles and neutrons, we obtain the conservation law for $n_B$,

\begin{equation}
\label{densb}
\frac{\partial n_B}{\partial t}+\vec{\nabla}\cdot(n_n\vec{v_n}+n_p\vec{v_p})=0.
\end{equation}

\section{One-Dimensional Model}

In this section, we consider a one-dimensional model where the magnetic field in the neutron star points in one Cartesian direction but varies only along an orthogonal direction as $\vec{B}(\vec{r},t)=B_z(x,t)\hat{z}$. We take this as a perturbation on a non-magnetized background in hydrostatic and chemical equilibrium. The physical variables characterizing the background are labeled by the sub-index $0$. We perturb this background by introducing a magnetic field, which causes small perturbations of the particle densities. Thus, we write the particle densities as $n_i(x,t)= n_{0i}(x)+\delta n_i(x,t)$, where $i=e,p,n$, and $\delta n_i(x,t)\ll n_{0i}(x)$. The chemical potentials are $\mu_i(x,t)=\mu_{0i}(x)+\delta \mu_i(x,t)$, where $\delta \mu_i(x,t)=k_{iB}\delta n_B(x,t)+k_{ic}\delta n_c(x,t)$, $k_{iB}=\left(\partial \mu_{i}/\partial n_{B}\right)_{n_{0c}}$, and $k_{ic}=\left(\partial \mu_{i}/\partial n_{c}\right)_{n_{0B}}$, with $k_{eB}=0$. We use the Cowling approximation, thus, we neglect the perturbations of the gravitational potential respect to the background value, i.e., $\psi=\psi_0$. From Ampere's law, $j_x=(c/4\pi)(\vec{\nabla}\times\vec{B})_x=0=n_ce(v_{px}-v_{ex})$, thus, $v_{ex}=v_{px}=v_{cx}$. Using Eqs.~\ref{magn}, \ref{ambi}, \ref{Hall}, \ref{densn}, and \ref{densb}, we obtain the following dimensionless set of equations for the evolution of the magnetic field and the particles:

\begin{equation}
\label{magbar}
\frac{\partial \overline{B}_z}{\partial \overline{t}}=-\overline{n}_{cs} \overline{\Upsilon}\frac{\partial (\overline{v}_{cx}\overline{B}_z)}{\partial \overline{x}}+\overline{\omega}\frac{\partial }{\partial \overline{x}}\left(\overline{r}_0 \frac{\partial \overline{B}_z}{\partial \overline{x}}\right),
\end{equation}

\begin{equation}
\label{nbnorm}
\frac{\partial \delta \overline{n}_{B}}{\partial \overline{t}}= -\frac{\partial}{\partial \overline{x}}\left(\overline{n}_{0n} \overline{v}_{nx}+ \overline{n}_{0c}\overline{v}_{cx}\right),
\end{equation}

\begin{equation}
\label{ncnorm}
\frac{\partial \delta \overline{n}_{c}}{\partial \overline{t}}=-\frac{\partial}{\partial \overline{x}}\left(\overline{n}_{0c}\overline{v}_{cx}\right)-\overline{\lambda}\overline{n}_{cs}\overline{\vartheta}\left(\overline{k}_{0c} \delta \overline{n}_{c}+\overline{k}_{0B} \delta \overline{n}_{B}\right),
\end{equation}

where

\begin{eqnarray}
\label{vnbar}
\overline{v}_{nx}=-\overline{\mu}_{0n}\frac{\partial}{\partial \overline x}\left(\frac{\overline{k}_{nB}\delta \overline{n}_B+\overline{k}_{nc}\delta \overline{n}_c}{\overline{\mu}_{0n}}\right)-\frac{\overline{n}_{0c}\overline{\mu}_{0n}}{\overline{n}_{0n}}\frac{\partial}{\partial \overline x}\left(\frac{\overline{k}_{cc}\delta \overline{n}_c+\overline{k}_{pB}\delta \overline{n}_B}{\overline{\mu}_{0n}}\right)\nonumber\\-\frac{1}{\overline{n}_{0n}}\frac{\partial(\overline{B}_z^2)}{\partial \overline{x}} ,
\end{eqnarray}

\begin{eqnarray}
\label{vabar}
\overline{v}_{Ax}=\frac{\overline{\epsilon}}{\overline{n}_{0c}\overline{\gamma}_{cn}}\left[-\frac{\overline{n}_{0c}\overline{\mu}_{0n}}{\overline{n}_{0n}}\frac{\partial}{\partial \overline x}\left(\frac{\overline{k}_{cc}\delta \overline{n}_c+\overline{k}_{pB}\delta \overline{n}_B}{\overline{\mu}_{0n}}\right)-\frac{1}{\overline{n}_{0n}}\frac{\partial(\overline{B}_z^2)}{\partial \overline{x}}\right].
\end{eqnarray}

In the above equations we have defined dimensionless variables as $\overline{a}=a/a_s$, with the sub-index $s$ standing for a characteristic value of the corresponding dimensional variable. We set the characteristic values as: $x_s=L$, $B_s=B_{z}^{max}$, $t_s= \alpha x_s^2/(n_{s}k_{s})$, $v_s=B_s^2/(8\pi n_s x_s \alpha)$, $\delta n_s=B_s^2/(8\pi n_s k_s)$, $n_s=n_{0n}^{max}$, $n_{cs}=n_{0c}^{max}$, $k_s=k_{nB}^{max}$, $\mu_s=\mu_{0n}^{max}$, $r_s=r_0^{max}$, $\gamma_s=\gamma_{cn}^{max}=\gamma_{pn}^{max}+\gamma_{en}^{max}$. On the other hand, we have defined $r_0=c^2/4\pi\sigma_0$, $\overline{k}_{0c}=\overline{k}_{cc}+\overline{k}_{nc}$, $\overline{k}_{cc}=\overline{k}_{pc}+\overline{k}_{ec}$, $\overline{k}_{0B}=\overline{k}_{pB}-\overline{k}_{nB}$, $\overline{n}_{cs}=n_{cs}/n_{s}$. The dimensionless parameters $\overline \Upsilon \equiv B_s^2/(8\pi n_{cs}n_{s}k_{s})$, $\overline{\epsilon} \equiv \alpha/(n_{s}\gamma_{s})$, $\overline{\vartheta}\equiv \alpha \lambda_s x_s^2/n_{cs}$, and  $\overline{\omega}\equiv \alpha r_{s}/(n_{s}k_{s})$ define the importance of the different processes.

\section{Results}

\subsection{Characteristic evolutionary time scales}

The Eqs.~\ref{magbar}, \ref{nbnorm}, \ref{ncnorm} describe the temporal evolution of the variables $\overline{B}_z$, $\delta \overline{n}_{B}$, and $\delta \overline{n}_{c}$ respectively. The characteristic time scales on which this evolution occurs can be estimated from these equations. The two terms on the right hand side (RHS) of Eq.~\ref{magbar} give the characteristic time scale on which the magnetic field evolves due to the coupling with the particles and due to ohmic diffusion. On the other hand, the RHS of Eq.~\ref{nbnorm} gives the time scale on which a baryonic density perturbation evolves due to the motion of the particles. Finally, the two terms on the RHS of Eq.~\ref{ncnorm} give the characteristic time scale on which a charged density perturbation evolves due to the motion of the particles and due to weak interactions between them. In the next paragraphs we give an order-of-magnitude estimative of each one of these characteristic time scales and describe the quasi-equilibria that are achieved along the evolution.     

\subsubsection{Time scale to achieve magneto-hydrostatic quasi-equilibrium}

When a magnetic field is present in the system, the magnetic force pushes on the charged particles. At the first stages of the evolution, the collisional coupling between the charged particles and neutrons makes them move with about the same velocity, thus, $v_{cx}\approx v_{nx}$. Neglecting weak interactions, from Eqs.~\ref{nbnorm}, \ref{ncnorm} we obtain $\delta n_B / n_{0B} \approx \delta n_c/n_{0c} \ll 1$. The induced charged-particle and neutron pressure gradients tend to choke the magnetic force. The magneto-hydrostatic quasi-equilibrium is reached when there is a close balance between these oposing forces (see RHS of Eq.~\ref{vnbar}). In the following, we estimate the time scale over which this close-balance occurs. Considering that $n_{0c}\ll n_{0n}$, from Eq.~\ref{vnbar} we obtain the associated velocity to a given perturbation $\delta {n}_B$ as $v_{nx}\sim v_{cx}\sim (1/\alpha L)\left[k_{nB}+(n_{0c}/n_{0B})k_{nc}\right]\delta n_B$. From the Eq.~\ref{nbnorm}, we estimate the time-scale to produce this perturbation as

\begin{equation}
\label{tnn}
t_{{\delta n}_B}\sim \frac{L\delta n_B}{n_{0B}v_{nx}}\sim \frac{x_s\delta n_s}{n_{s}v_{s}}=\frac{\alpha x_s^2}{n_{s}k_{s}}=t_s.
\end{equation}

We require this time scale to be the shortest, so we define it as our fundamental unit of time. Note that the non-physical parameter $\alpha$ in Eq.~\ref{tnn} controls how fast the magneto-hydrodystatic equilibrium is reached.  

\subsubsection{Time scale for charged particles to reach diffusive quasi-equilibrium}

Now we suppose that the quasi-equilibrium discussed in the last section has been achieved (i.e., $v_{cx}=v_{nx}+v_{Ax}\sim v_{Ax}$). Thus, only the charged-particles pressure gradients balances the magnetic force. The charged-particle diffusive quasi-equilibrium is reached when there is a close balance between these oposing forces (see RHS of Eq.~\ref{vabar}). From Eq.~\ref{vabar}, we obtain the associated velocity to a given perturbation  $\delta {n}_c$ as $v_{Ax}\sim ((k_{cc}+k_{pB})\delta n_c)/(n_{0n}\gamma_{cn}L)$. Neglecting weak interactions, from Eqs.~\ref{nbnorm}, \ref{ncnorm}, and \ref{tnn} we estimate the time scale to produce this perturbation,

\begin{equation}
\label{bartnc}
\overline{t}_{{\delta n}_c}\sim \frac{n_s\gamma_{s}}{\alpha}= \frac{1}{\overline{\epsilon}}\gg 1.
\end{equation}

Note that the dimensionless parameter $\overline{\epsilon}$ is inversely proportional to the collisional strength parameter $\gamma_{cn}$ between the charged and neutral particles, thus $\gamma_{s}$ controls how fast the charged fluid reaches the diffusive equilibrium. On the other hand, we require $\overline \epsilon \ll 1 $, so $\overline{t}_{{\delta n}_c}\gg \overline{t}_s= 1$. 

\subsubsection{Time scale to achieve chemical equilibrium}
 
The characteristic time scale on which a chemical imbalance is erased by the weak interactions can be estimated from Eq.~\ref{ncnorm}. Neglecting the first term on the RHS of Eq.~\ref{ncnorm}, comparing the terms with $\delta n_c$, and using Eq.~\ref{tnn}, we obtain

\begin{equation}
\label{bartdmu}
\overline{t}_{\Delta \mu}\sim \frac{n_{cs}}{\alpha \lambda_s x_s^2}=\frac{1}{\overline{\vartheta}}\gg 1,
\end{equation}

We require $\overline \vartheta \ll 1 $, so $\overline{t}_{\Delta \mu}\gg \overline{t}_s=1$. Note that, in Eq.~\ref{bartdmu} the parameter $\lambda$ controls how fast the chemical equilibrium is achieved.

\subsubsection{Ohmic diffusion time scale}

The time scale on which the magnetic field evolves by ohmic diffusion can be estimated from Eq.~\ref{magbar}. Neglecting the first term on the RHS of Eq.~\ref{magbar} and using Eq.~\ref{tnn}, we get

\begin{equation}
\label{bartohm}
\overline{t}_{ohmic}\sim \frac{n_{s}k_{s}}{\alpha r_s} \equiv \frac{1}{\overline{\omega}}\gg 1.
\end{equation}

In Eq.~\ref{bartohm}, the parameter $\overline \omega$, which is directly proportional to the parameter $r_s$, controls how fast the magnetic field decays by the electrical resistivity. We require $\overline \omega \ll 1 $, so $\overline{t}_{ohmic}\gg \overline{t}_s =1$.
 
\subsubsection{Time scale for the evolution of the magnetic field ignoring ohmic diffusion}

Here we neglect the ohmic diffusion term in Eq.~\ref{magbar} and assume the system has reached the magnetohydrostatic equilibrium ($v_{nx}\sim 0$). In order to estimate the time scale for the evolution of the magnetic field, we have to distinguish between two cases. In the first case $t_{\delta n_c} \ll t_{\Delta \mu}$, and we want to describe the evolution of the magnetic field for times longer than $t_{\delta n_c}$, thus, the charged particles will achieve diffusive equilibrium before chemical equilibrium is restored. The necessary ambipolar velocity to achieve a quasi-equilibrium state on which the charged density perturbations do not change can be estimated from Eq.~\ref{ncnorm} as $v_{Ax}\sim (\delta n_c L \lambda (k_{0c}+k_{0B}))/n_{0c}$. On the other hand, if we neglect the ohmic resistivity, from Eq.~\ref{magbar} we can estimate the time scale $L/v_{Ax}$ on which the magnetic field evolves as

\begin{equation}
\label{tb2}
\overline{t}_B \sim \frac{\overline {t}_{\Delta \mu}}{\overline \Upsilon}\sim \frac{1}{\overline\Upsilon~ \overline \vartheta}.
\end{equation}

The opposite case corresponds to  $t_{\delta n_c} \gg t_{\Delta \mu}$ and we want to describe the evolution for times longer than $t_{\Delta \mu}$. In this case, the system achieves chemical equilibrium before the charged particles achieve diffusive equilibrium, thus during this time scale $v_{Ax}\neq 0$. From Eq.~\ref{magbar}, $t_B\sim L/v_{Ax}$ and from Eq.~\ref{vabar} we have, $v_{Ax}\sim B_z^2/(n_{0n}n_{0c}\gamma_{cn}8\pi L)$. From Eq.~\ref{magbar} we can estimate the time scale for the evolution of the magnetic field as

\begin{equation}
\label{tambapr}
\overline{t}_{B}\sim \frac{\overline{t}_{{\delta n}_c}}{\overline \Upsilon} \sim \frac{1}{\overline\Upsilon~ \overline \epsilon}.
\end{equation}

\subsection{Evolution of a Linear Magnetic Field Perturbation}

In this section, we study the evolution of a linear magnetic field perturbation $\delta \overline{B}_z$. We write the magnetic field as: $\overline{B}_z(\overline x, \overline t)=1+\delta \overline{B}_z(\overline x, \overline t)$, where $\delta B_z(\overline x, \overline t)\ll1$. We start the simulation with an initial magnetic field perturbation of the form $\delta \overline B_z(\overline x, 0)=\overline A \cos(l\pi\overline{x})$, and a homogeneous particle background: $\delta \overline{n}_c(\overline x, 0)/\overline{n}_{0c}=\delta \overline{n}_n(\overline x, 0)/\overline{n}_{0n}=0$. For this perturbation, the characteristic length on which the magnetic force varies is $L/(l\pi)$, thus, $\overline {t}_s=1/(l\pi)^2= 0.025$. We define $n\equiv\overline{n}_{0c}$, $\overline{k}_{nc}\equiv k$, $\varrho\equiv n\overline{k}_{cc}$, $\beta\equiv \overline{n}_{0c}\overline{k}_{pB}$. In the interior of a neutron star we have the typical values $n_{0n}=10^{38} cm^{-3}$, $n=3.9\times 10^{-2}$, $\beta=-0.0775$, $k=-2.962$ and $\rho=1.508$ \cite{A-98}. In our analysis, we neglect the ohmic diffusion ($\omega=0$). We use the parameters: $\overline \omega=0$, $\overline \Upsilon=\overline \vartheta=0.05,  \overline \epsilon =0.005$,  $\overline A=-0.0001$, and $l=2$. For these parameters, $\overline{t}_{{\delta n}_c}=5$ and $\overline{t}_{\Delta \mu}=20$. The characteristic time scale to reach the diffusive charged-fluid quasi-equilibrium can be estimated as the minimun between these two values. Since $\overline{t}_{{\delta n}_c}\ll \overline{t}_{\Delta \mu}$, the magnetic field significantly evolves on the time scale, $\overline{t}_B\sim 1/(\overline{\Upsilon}~\overline{\vartheta})=400$. In Fig. 1, we show the system's evolution during the time scale to achieve magneto-hydrostatic equilibrium. We label the different instants of this evolution with progressive numbers, i.e., the initial condition is labeled with the number ($1$). We see from Fig. 1 that at the end of this time scale, $\delta \overline{n}_c/\overline{n}_{0c}\sim\delta \overline{n}_n/\overline{n}_{0n}$, and the magnetic perturbation has not evolved significantly. In Fig. 2, we show the system's evolution during the time scale where the diffusive charged quasi-equilibrium is reached. The plots labeled with number $1$ correspond to the final instant of the previous scale, i.e., the scale where the system achieved the magneto-hydrostatic equilibrium. We see in Fig. 2 that at the end of this time scale, $\delta \overline{n}_n/\overline{n}_{0n}\sim 0$ and $\delta \overline{n}_c/\overline{n}_{0c}$ has significantly grown in order to balance the magnetic pressure. The magnetic perturbation has begun to evolve significantly. It is also shown in Fig. 2, that at the end of this time scale, the separation between the lines is narrower, which shows that the charged diffusive quasi-equilibrium has been reached. In Fig. 3 we show the system's evolution during the time scale where the chemical equilibrium is restored and the magnetic perturbation decays. Again, the plots labeled with number $1$ correspond to the final instant of the previous scale, i.e., the scale where the system achieved the charged diffusive quasi-equilibrium. We see in Fig. 3 that at the end of this time scale the density perturbations have been erased and the magnetic field perturbation has decayed.

\section{Conclusions}

Using numerical simulations in one dimension, we studied the effects of some non-ideal MHD processes on the magnetic field evolution in neutron stars. We found that the system evolves through succesive quasi-equilibria, and we estimated the characteristic time scales on which these quasi-equilibria occur.

\begin{center}
Acknowledgments
\end{center}
We acknowledge financial support through FONDECYT postdoctoral project 3060103, Gemini project 32070014, and regular FONDECYT projects 1060644 and 1070854. We also thank the ESO-Chile Mixed Committee.

\begin{figure}[h]
\includegraphics[height=4cm]{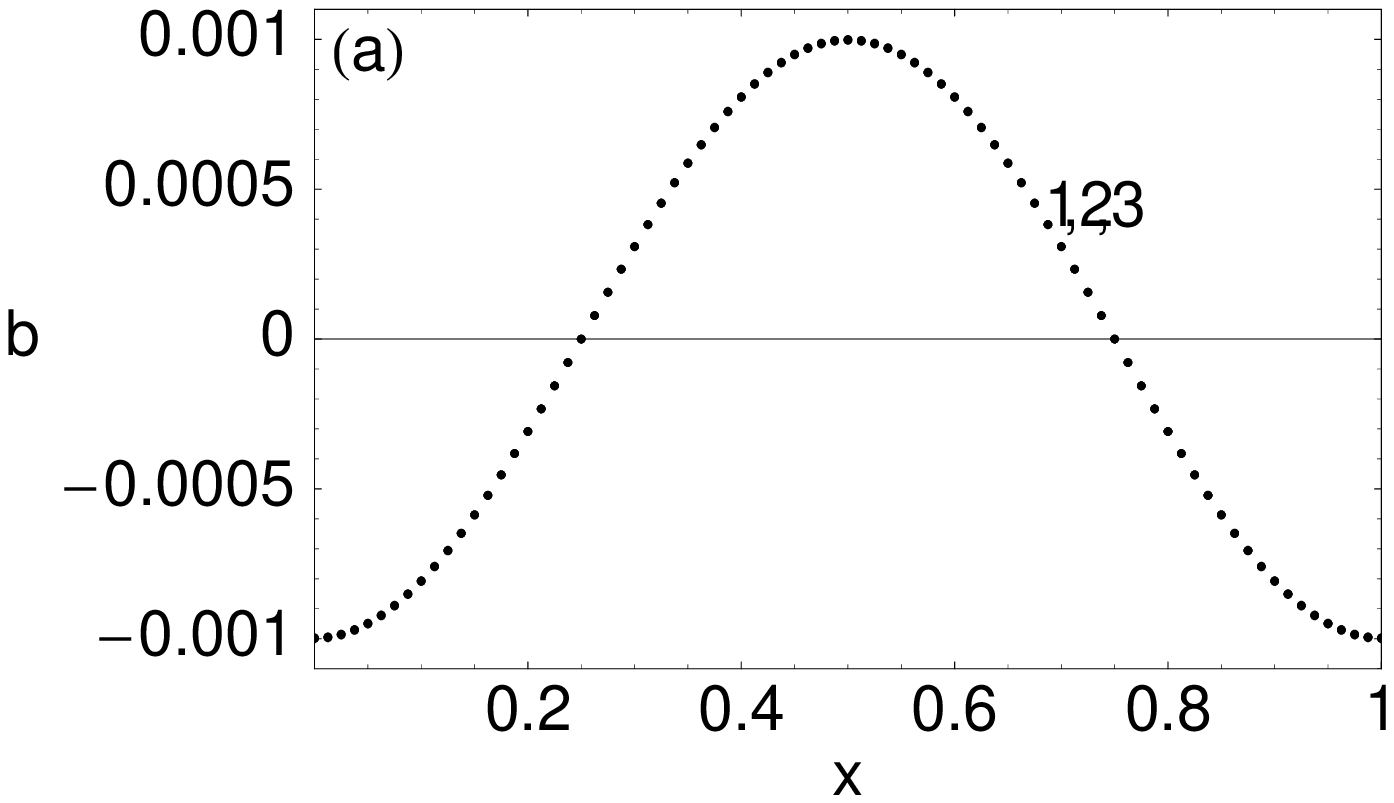}
\includegraphics[height=4cm]{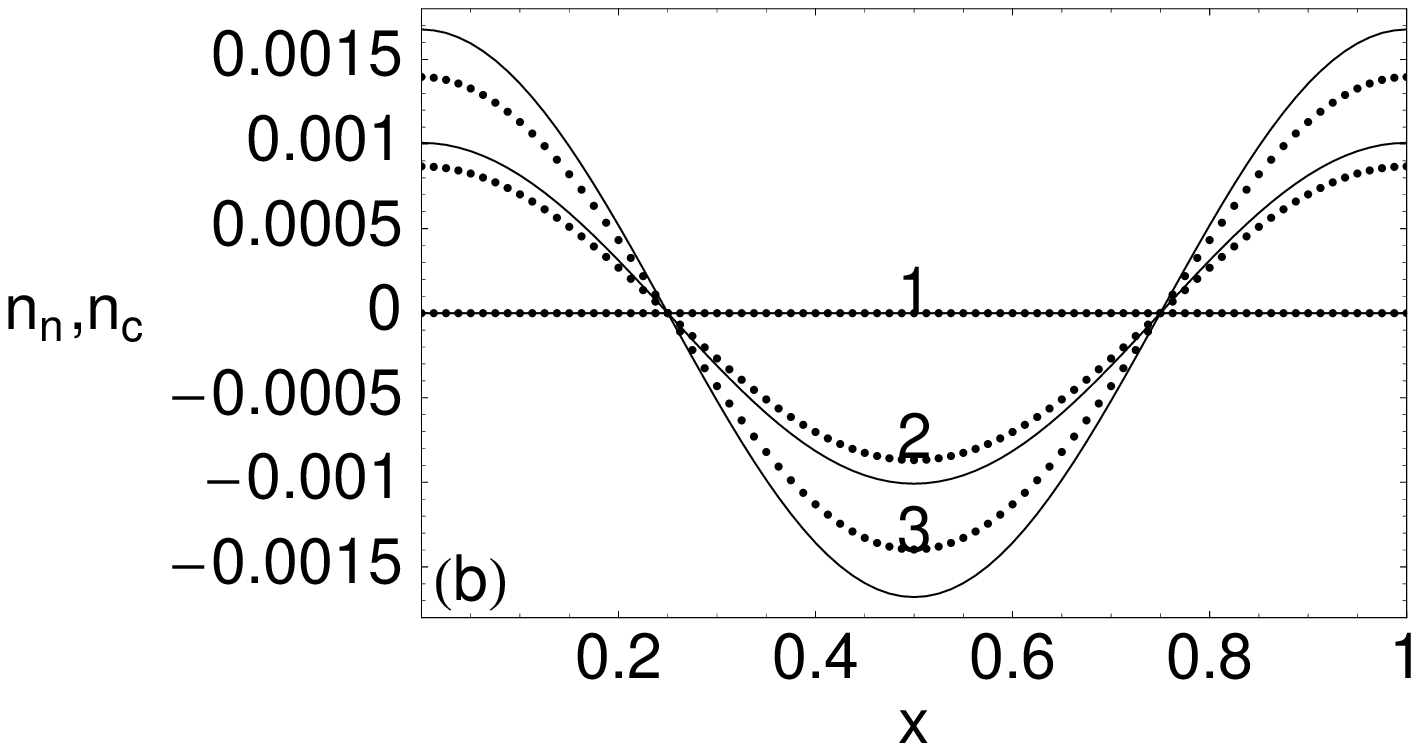}
\caption{Magneto-Hydrostatic quasi-equilibrium. We have defined, $b\equiv\delta \overline{B}_z$, $n_b\equiv \delta \overline{n}_c/\overline{n}_{0c}$, $n_n\equiv \delta \overline{n}_n/\overline{n}_{0n}$. Dotted lines in Fig. 1b. refer to the neutrons and full lines to charged particles. Label times are: (1): $\overline t=0 $, (2): $\overline t=0.013986 $,  (3): $\overline t=0.0279725$.}
\end{figure}

\begin{figure}[h]
\includegraphics[height=4cm]{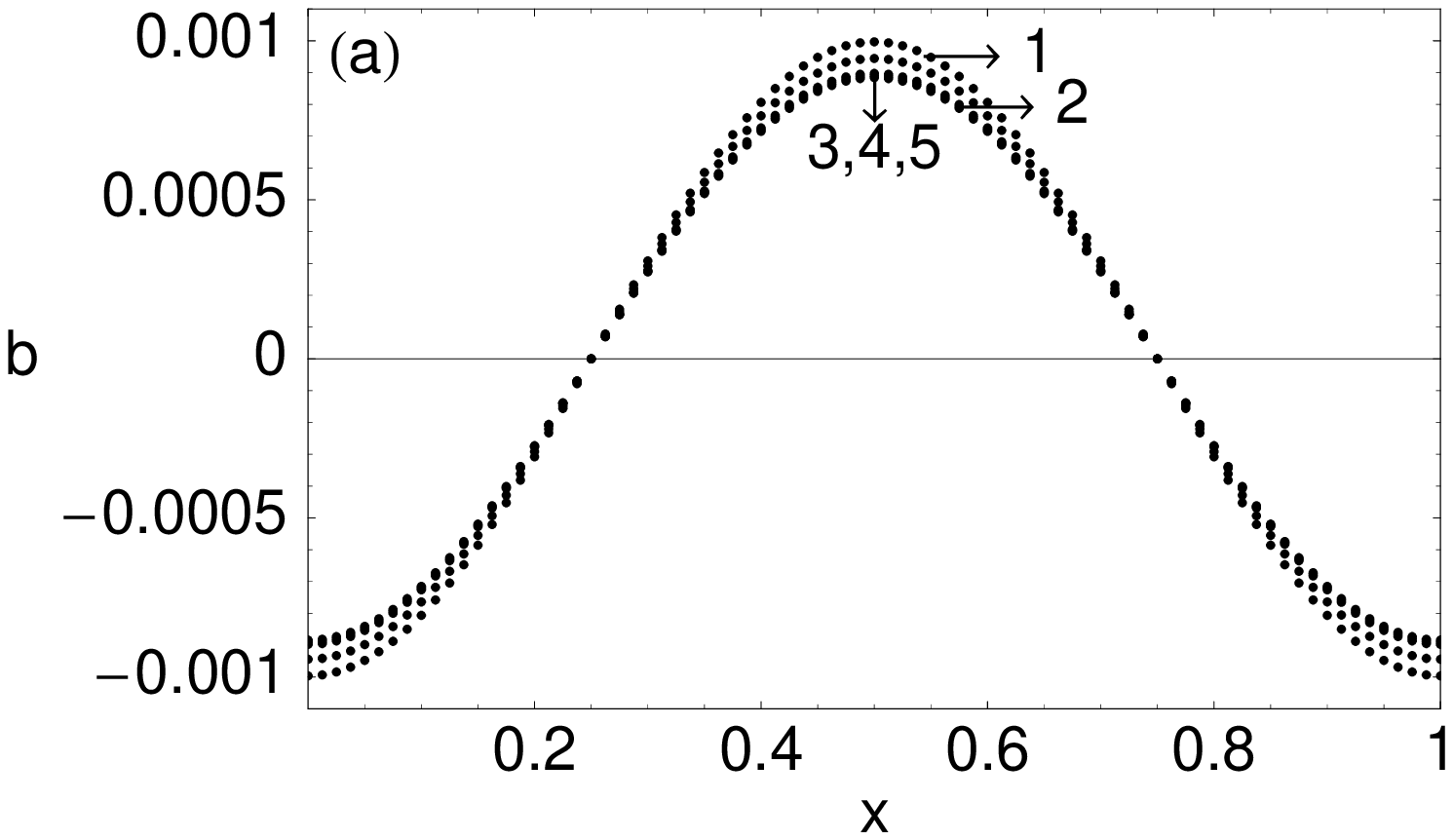}
\includegraphics[height=4cm]{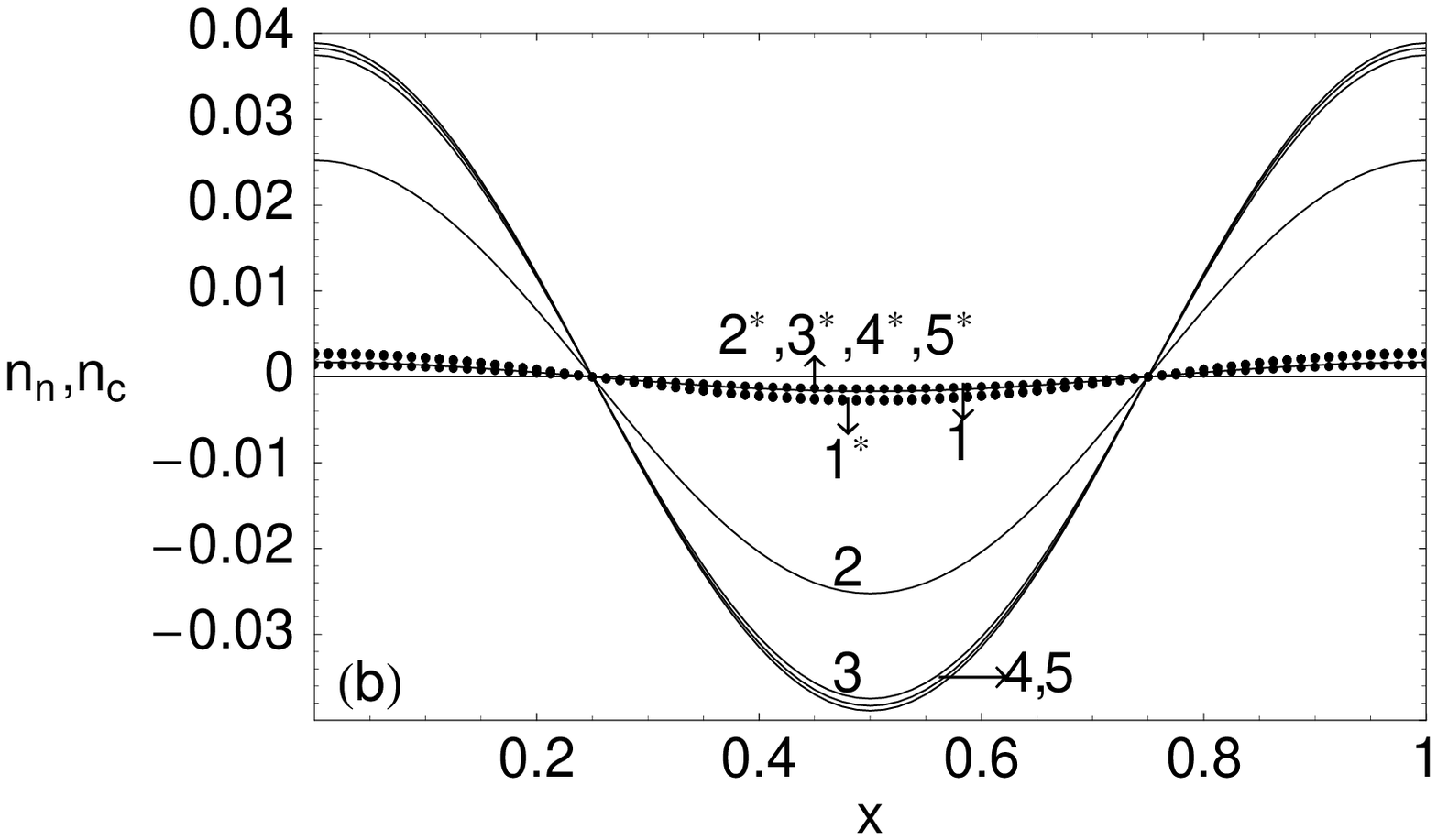}
\caption{Diffusive Charged Fluid Quasi-Equilibrium. We use the same conventions than Fig. 1. The label times are: (1,$1^*$): $\overline t=0.0279725$, (2, $2^*$): $\overline t=3.23304$,  (3, $3^*$): $\overline t=8.57481$, (4, $4^*$): $\overline t=9.64316$, (5, $5^*$): $\overline t=10.7115$. The upper asteristik in each number refers to neutrons.}
\end{figure}

\begin{figure}[h]
\includegraphics[height=4cm]{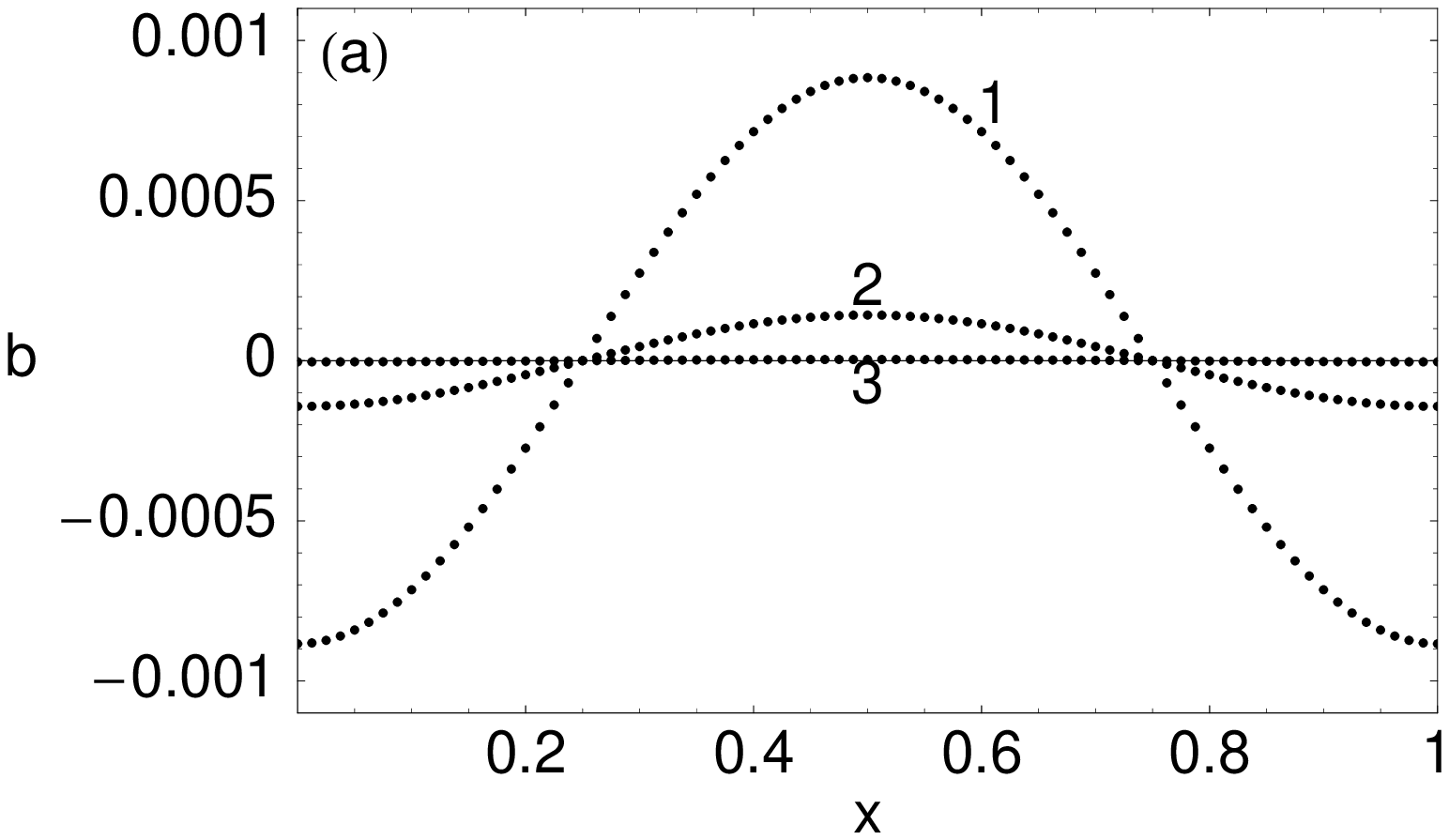}
\includegraphics[height=4cm]{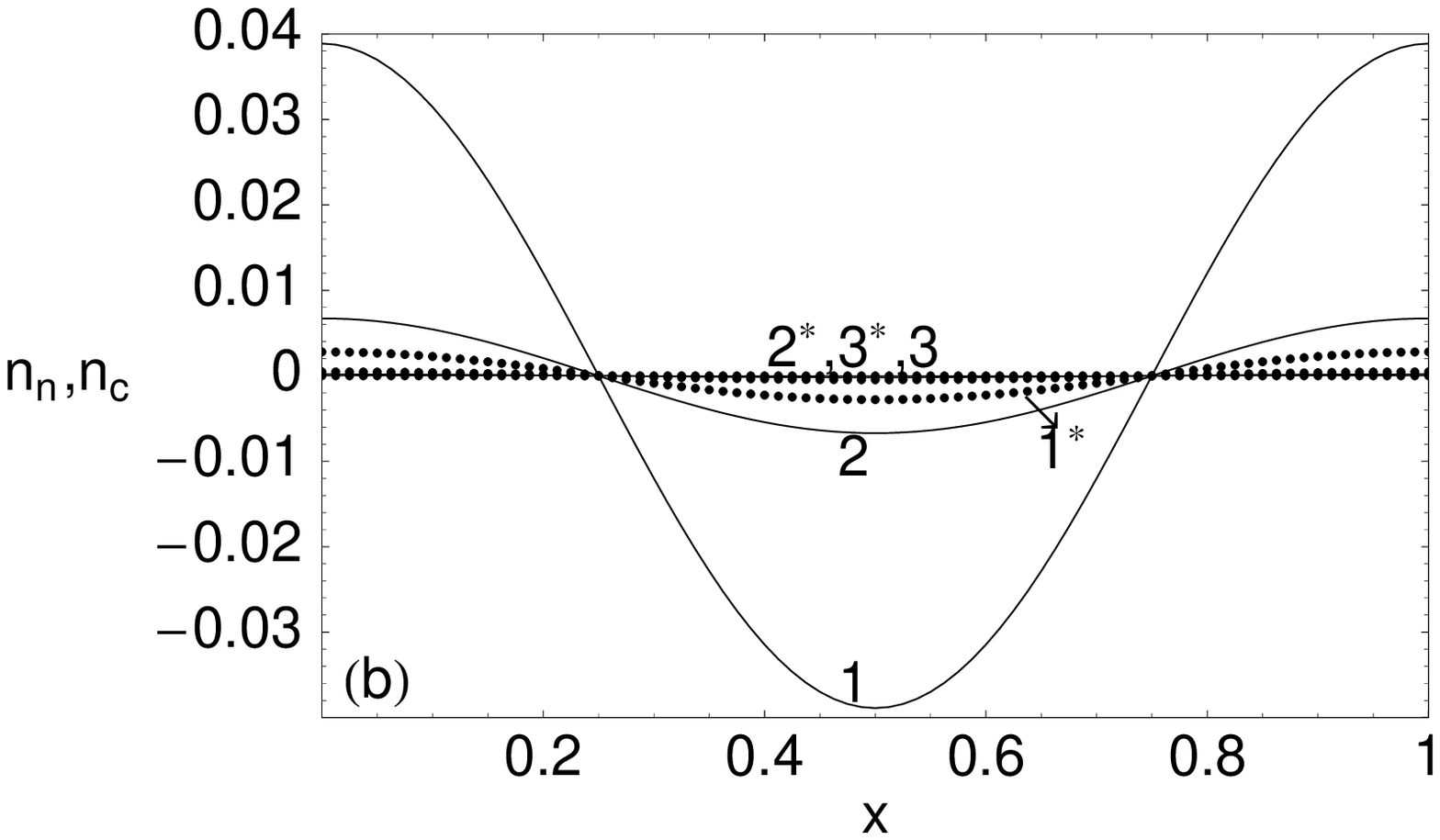}
\caption{Decay of the magnetic field and Chemical Equilibrium. We use the same conventions than Fig. 1 and Fig.2. The label times are: (1,$1^*$): $\overline t=10.7115$, (2, $2^*$): $\overline t=116.83$, (3, $3^*$): $\overline t=329.067$.}
\end{figure}

\end{document}